\RequirePackage{lineno}
\documentclass[twocolumn,amsmath,amssymb,showpacs,superscriptaddress,unsortedaddress, nofootinbib]{revtex4}
\usepackage{epsf}
\usepackage{graphicx}
\usepackage{siunitx}
\usepackage{multirow}
\usepackage{color}
\usepackage{soul}
\usepackage{subfigure}
\usepackage[absolute,overlay]{textpos}

\begin{document}

\title{Effect of electron doping in FeTe$_{1-y}$Se$_{y}$ realized by Co and Ni substitution}

\author{M. Rosmus}
\affiliation{M. Smoluchowski Institute of Physics, Jagiellonian University, {\L}ojasiewicza 11, 30-348, Krak\'{o}w, Poland}

\author{R. Kurleto}
\affiliation{M. Smoluchowski Institute of Physics, Jagiellonian University, {\L}ojasiewicza 11, 30-348, Krak\'{o}w, Poland}

\author{D. J. Gawryluk }
\email[Corresponding author: ]{gawryluk@ifpan.edu.pl}
\affiliation{Laboratory for Multiscale Materials Experiments, Paul Scherrer Institut, 5232 Villigen PSI, Switzerland}
\affiliation{Institute of Physics, Polish Academy of Sciences,  Lotnik\'{o}w 32/46, 02-668 Warszawa, Poland}

\author{J. Goraus}
\affiliation{Institute of Physics, University of Silesia, ul. 75 Pu{\l}ku Piechoty 1A, 41-500 Chorz\'{o}w, Poland}

\author{M. Z. Cieplak }
\affiliation{Institute of Physics, Polish Academy of Sciences,  Lotnik\'{o}w 32/46, 02-668 Warszawa, Poland}

\author{P. Starowicz}
\email[Corresponding author: ]{pawel.starowicz@uj.edu.pl}
\affiliation{M. Smoluchowski Institute of Physics, Jagiellonian University, {\L}ojasiewicza 11, 30-348, Krak\'{o}w, Poland}

\begin{textblock*}{10cm}(9.75cm,26.5cm) 

 \textit{Preprint submitted to Superconductor Science and Technology}

\end{textblock*}

\date{\today}
\begin{abstract}

Angle-resolved photoemission spectroscopy (ARPES) reveals effects of electron doping, which is realized by Co and Ni substitution for Fe in FeTe$_{1-y}$Se$_{y}$ (y$\sim$0.35) superconductor. The data show consistent band shifts as well as expansion and shrinking of electron and hole Fermi surface, respectively. Doping of either element leads to a Lifshitz transition realized as a removal of one or two hole pockets. This explains qualitatively a complex behavior of Hall coefficient observed before [Bezusyy, et al., Phys. Rev. B \textbf{91}, 100502 (2015)], including change of sign with doping, which takes place only below room temperature. Assuming that Ni substitution should deliver twice more electrons to the valence band than Co, it appears that such transfer is slightly more effective in the case of Co. Therefore, charge doping cannot account for much stronger effect of Ni on superconducting and transport properties [Bezusyy, et al., Phys. Rev. B \textbf{91}, 100502 (2015)]. Although overall band shifts are roughly proportional to the amount of dopant, clear deviations from a rigid band shift scenario are found. The shape of electron pockets becomes elliptical only for Ni doping, effective mass of electron bands increases with doping, strong reduction of effective mass is observed for one of hole bands of the undoped system. The topology of hole and electron pockets for superconducting Fe$_{1.01}$Te$_{0.67}$Se$_{0.33}$ with T$_{c}$=13.6~K indicates a deviation from nesting. Co and Ni doping causes further departure from nesting, which accompanies the reduction of critical temperature. 
\end{abstract}


\maketitle

\section{Introduction}

The high critical temperature and evidence of unconventional pairing mechanism attracted attention of many researchers to iron based superconductors.\cite{Kamihara2008, Stewart2010, XChen2014, QSi2016} Although spin fluctuations are considered to be responsible for this mechanism, an important goal is to settle physical properties of a given compound, which are real prerequisites of superconductivity. For example, topology of Fermi surface (FS) corresponding to optimal doping does not seem to be universal for different superconducting families.\cite{Ye2014} Nevertheless, marginal breakdown of nesting between hole and electron pockets may be advantageous for superconductivity, as this was proposed to support spin fluctuation mechanism.\cite{Lee2012}

Chemical doping remains the key method to reveal phase diagrams of these systems with superconducting phase located typically in a vicinity of magnetic order. A realization of this can be accomplished e.g. by a substitution of transition metals for iron. Among different elements Co and Ni as dopants lead to quite successful suppression of magnetic order and induction of superconductivity in 122\cite{Sefat2008PRL, LeitheJasper2008, LiNJP2009} and 1111\cite{Sefat2008PRB, Wang2009, Cao2009} iron pnictides. In contrast, these elements suppress superconductivity in highly correlated 11 iron chalcogenides.\cite{Shipra2010, Gawryluk2011, Bezusyy2015, Kumar2012, Nabeshima2012, Zhang2013, Inabe2013} The real effect of substituted Co or Ni in both iron pnictides and chalcogenides is a subject of debate. According to a number of theoretical predictions \cite{Wadati2010,Kemper2009, Berlijn2012} and experimental results \cite{Zhang2013, Bittar2011, Levy2012} these mainly act as scattering centers, which broaden bands and destroy magnetic order, allowing of superconductivity. In this scenario the effect of charge doping might be secondary or even quite negligible. On the other hand, angle-resolved photoemission spectroscopy (ARPES) studies \cite{Sekiba2009, Brouet2009, Liu2010, Liu2011, Neupane2011, Ideta2013, Vilmercati2016, Malaeb2009, Miao2015, Thirupathaiah2016} indicate the effect of electron doping, which is more efficient in the case of Co as compared to Ni.\cite{Ideta2013} Quite ambivalent conclusions support scenarios of either rigid \cite{Neupane2011, Ideta2013} or non-rigid \cite{Vilmercati2016, Thirupathaiah2016} band shift, in the latter case dopants influence correlations. A non-rigid band shift is reported in the first paper devoted to ARPES studies of Ni doped FeTe$_{1-y}$Se$_{y}$. \cite{Thirupathaiah2016} ARPES data for Co doped FeTe$_{1-y}$Se$_{y}$ have not been published so far. It should be mentioned that FeTe$_{1-y}$Se$_{y}$ is of particular interest since the discovery of topologically non-trivial states in this system.\cite{PZhangScience2018}

It is known that both Ni and Co doping suppresses superconductivity in FeTe$_{1-y}$Se$_{y}$ chalcogenides.\cite{Shipra2010, Gawryluk2011, Bezusyy2015, Kumar2012, Nabeshima2012, Zhang2013, Inabe2013} Systematic investigations of transport and magnetic properties were performed for Co and Ni substituted FeTe$_{1-y}$Se$_{y}$ (y$\sim$0.35).\cite{Bezusyy2015} It was found that the superconducting critical temperature, which equals 14~K for the undoped material, is reduced with both dopants. The reduction rate per doped electron appears to be much faster in case of Ni, considering that this element supplies twice more electrons than Co. Moreover, nickel increases considerably electrical resistivity and is considered to be a source of more efficient impurity scattering and correlations. Finally, the Hall coefficient (R$_{H}$) exhibits a complex behavior analogical for both dopants. At high temperature it remains positive for all dopant concentration, whereas at low temperature it increases first and then becomes negative showing another nonmonotonic behavior with a minimum. These are symptoms of interesting modifications of the band structure and FS.

The complex behavior found in the transport studies \cite{Bezusyy2015} motivated us to investigate the influence of Co and Ni doping on the electronic structure of FeTe$_{1-y}$Se$_{y}$ (y$\sim$0.35) by means of ARPES. The data indicate a clear effect of electron doping on both electron and hole pockets with Lifshitz transition observed in the hole FS. Although, energy shifts of different bands are comparable, some details indicate deviations from rigid band shift scenario. The data are able to explain the peculiar behavior of R$_{H}$.

\section{Experimental}
Single crystals of Fe$_{1-x}$M$_{x}$Te$_{1-y}$Se$_{y}$ (M=~Co or Ni, y$\sim$0.35) were grown using Bridgman’s method.\cite{Gawryluk2011} They were characterized by means of scanning electron microscopy with energy-dispersive X-ray spectroscopy (SEM-EDX), X-ray diffraction, AC magnetic susceptibility, measurements of electrical resistivity and R$_{H}$ as described in the supplementary material. \cite{Gawryluk2011, Bezusyy2015} Electronic structure was investigated for superconducting Fe$_{1.01}$Te$_{0.67}$Se$_{0.33}$ with a critical temperature of T$_{c}$=13.6~K as well as for one Co doped Fe$_{0.94}$Co$_{0.09}$Te$_{0.67}$Se$_{0.33}$ and two Ni doped: Fe$_{0.97}$Ni$_{0.05}$Te$_{0.65}$Se$_{0.35}$ and Fe$_{0.91}$Ni$_{0.11}$Te$_{0.65}$Se$_{0.35}$ systems. The Co doped crystal exhibited only an onset of the transition to superconductivity in resistivity (T$_{c,onset}$=4.6 K) and magnetic susceptibility (T$_{c,onset}$=9.5 K) without reaching zero resistivity down to 2 K (supplementary material). The Ni doped crystals were non-superconducting.  \footnote[6]{ARPES measurements have been performed for Fe$_{1-x}$M$_{x}$Te$_{1-y}$Se$_{y}$ (M = Co or Ni, y$\sim$0.35) crystals obtained with slow cooling method. In contrast, the data presented by Bezusyy et al. \cite{Bezusyy2015} were obtained for fast cooled crystals, which exhibited weaker crystal quality but better superconducting properties. Nevertheless, the effect of doping on electronic structure is comparable for different final stages of the preparation process.}

ARPES studies have been performed with Scienta R4000 hemispherical analyzer and He-I radiation ($h\nu=21.218$~eV). The energy resolution was set to 15~meV. Experimental setup is presented in Fig.~\ref{fig1}(a).  Partially polarized light from the monochromator (85\% of polarization) assures domination of $\sigma$-polarization for normal emission, where sample surface is perpendicular to a mirror plane. The measurements were conducted typically at the temperature of 18~K, which corresponds to non-superconducting state.

Density functional theory (DFT) calculations were performed with APW+lo method using ELK-LAPW code (version~4.3.06).\cite{Elk} Full relativistic approach and Perdew-Wang exchange-correlation potential  were employed. The accuracy of calculations was set to high-quality preset (rgkmax=8) and the number of k-points in the irreducible wedge of Brillouin zone was 50 (7x7x5 k-grid), or 10$^{6}$ (100x100x100 k-grid) for FS, which was found sufficient to give appropriate results.

\section{Results and discussion}
\begin{figure*}[!]
\includegraphics[width=\textwidth]{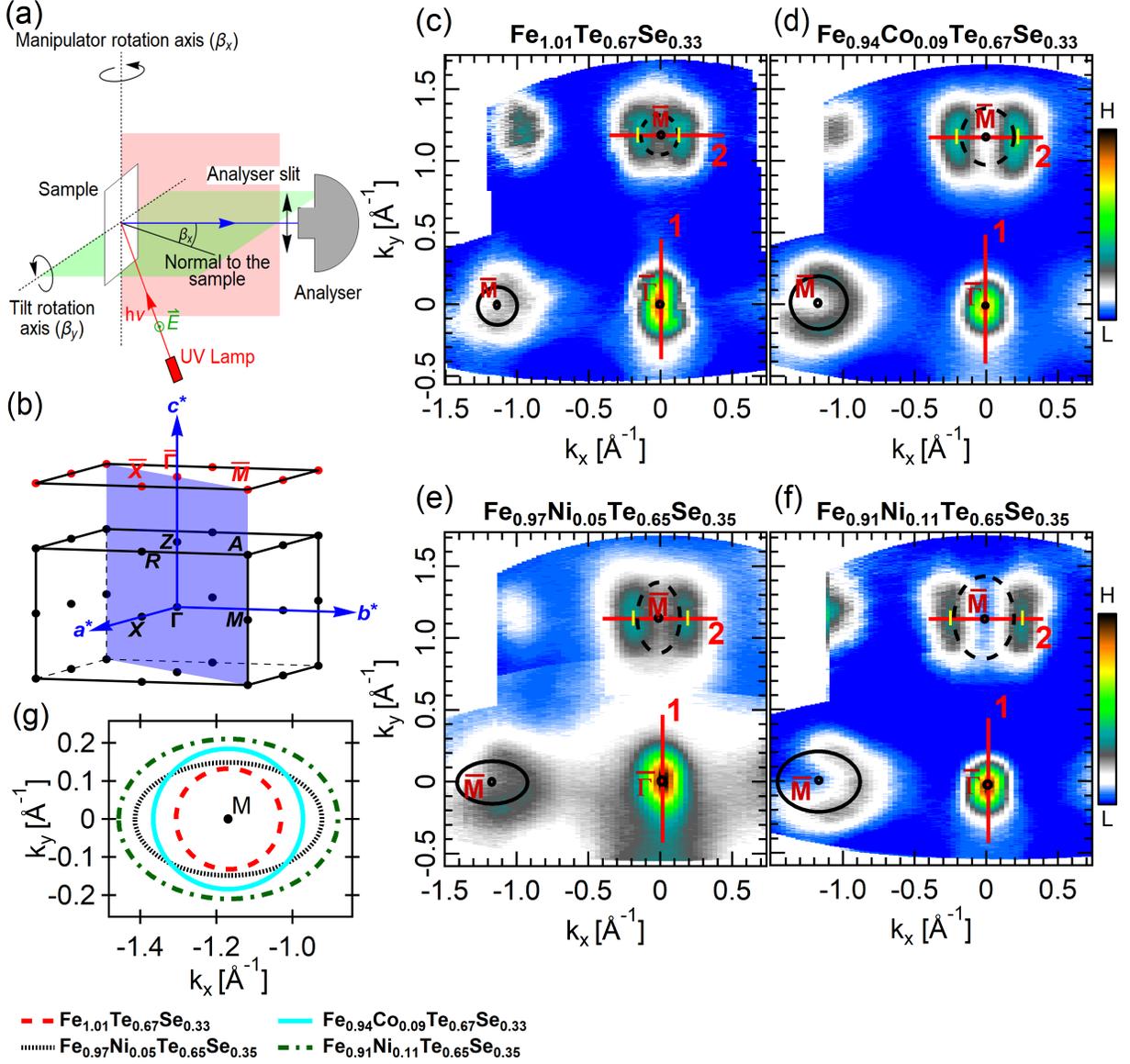}
\caption{Fermi surface mapping of Fe$_{1-x}$M$_{x}$Te$_{1-y}$Se$_{y}$ (M=Co, Ni, y$\sim$0.35) systems performed by means of angle-resolved photoemission spectroscopy (ARPES) using He-I ($h\nu=21.218$~eV) radiation at temperature of 18~K. Convention of 2 Fe atoms in elementary unit cell is used. (a) experimental geometry, (b) the first Brillouin zone for tetragonal FeTe structure and surface Brillouin zone with high symmetry points. Fermi surface scans are performed for (c) Fe$_{1.01}$Te$_{0.67}$Se$_{0.33}$, (d) Fe$_{0.94}$Co$_{0.09}$Te$_{0.67}$Se$_{0.33}$, (e) Fe$_{0.97}$Ni$_{0.05}$Te$_{0.65}$Se$_{0.35}$, (f) Fe$_{0.91}$Ni$_{0.11}$Te$_{0.65}$Se$_{0.35}$. Solid black ovals represent fitted electron pockets and dashed black lines are their replica obtained by symmetry operation. Solid red lines marked as~"1" and "2" denote paths along which ARPES spectra were taken for further analysis. Yellow markers drawn on both sides of M points at k$_{x}$=0 indicate Fermi vectors for the dispersion obtained along line~"2". (g) Contours of electron pockets obtained from momentum distribution  curve (MDC) fitting.}
\label{fig1}
\end{figure*}

FS mapping was performed for Fe$_{1.01}$Te$_{0.67}$Se$_{0.33}$ (referred to as 'undoped'  in the following), Fe$_{0.94}$Co$_{0.09}$Te$_{0.67}$Se$_{0.33}$, Fe$_{0.97}$Ni$_{0.05}$Te$_{0.65}$Se$_{0.35}$ and Fe$_{0.91}$Ni$_{0.11}$Te$_{0.65}$Se$_{0.35}$ systems by means of ARPES with the geometry shown in the Fig.~\ref{fig1}(a). We use a notation of two-dimensional (2D) surface Brillouin zone (Fig.~\ref{fig1}(b)) and a convention of two Fe atoms in the Brillouin zone. Hole and electron pockets are found for all the studied compositions near $\overline{\Gamma}$ and $\overline{M}$ points, respectively (Fig.~\ref{fig1}(c)-(f)). The expansion of electron pockets and shrinking of hole pockets (discussed further in the text) with Co and Ni doping is observed. Electron pockets reached from $\overline{\Gamma}$ by lowering k$_{x}$ (Fig.~\ref{fig1}(c)-(f)) have a circular shape for the undoped or Co doped system, whereas they become elliptical as a result of Ni doping. MDC fitting allowed to determine the shape of these pockets and the fits are drawn with solid lines at  k$_{x}\sim-1.2~\si{\angstrom}^{-1}$. We neglected the dispersion along c* and used the area of the elliptical fits to determine electron band filling, which was estimated as 2.3\%, 4.2\%, 3.8\% and 7.2\% for the undoped sample, Fe$_{0.94}$Co$_{0.09}$Te$_{0.67}$Se$_{0.33}$, Fe$_{0.97}$Ni$_{0.05}$Te$_{0.65}$Se$_{0.35}$ and Fe$_{0.91}$Ni$_{0.11}$Te$_{0.65}$Se$_{0.35}$, respectively. Hence, the effect of electron doping is well visible for the electron part of~FS. The presence of two electron pockets is not resolved and the given band filling corresponds to one of them. At this stage it is difficult to discuss the absence of the second pocket at k$_{x}\sim-1.2~\si{\angstrom}^{-1}$. Possibly, it is not seen due to unfavorable photoemission matrix elements. 

The pockets reached with increasing k$_{y}$ have a different shape. They exhibit a loss of intensity along k$_{x}=0$, which is due to a matrix element effect suppressing spectral intensity in this region of a k-space\cite{Brouet2012}. It becomes evident that dashed lines located around M point at k$_{y}\sim1.2 \si{\angstrom}^{-1}$, which represent replica of the pockets fitted at k$_{x}\sim-1.2~\si{\angstrom}^{-1}$, do not follow the spectra. There may be different explanations of this fact. First of all, a depletion of spectral weight at k$_{x}=0$ might influence the shape of the pockets determined from the ARPES experiment. However, dispersion obtained along the line “2” by MDC fitting does not converge with the dashed FS (For the dispersion along "2" the Fermi vectors are marked with yellow dashes in Fig.~\ref{fig1}(c)-(f).) This indicates that the pocket observed at k$_{y}\sim1.2~\si{\angstrom}^{-1}$ is different than that at k$_{x}\sim-1.2~\si{\angstrom}^{-1}$. This is not surprising, as it is known from DFT calculations and other ARPES results \cite{Thirupathaiah2016} that two electron pockets are present at M. Thus, a different shape of spectra at k$_{y}\sim1.2~\si{\angstrom}^{-1}$ and k$_{x}\sim-1.2~\si{\angstrom}^{-1}$ is related to different contribution from two electron pockets in these regions, which occurs due to different measurement geometry at k$_{y}\sim1.2~\si{\angstrom}^{-1}$ and k$_{x}\sim-1.2~\si{\angstrom}^{-1}$. We do not support the hypothesis of nonequivalence of band structure along “x” and “y” directions or the presence of orthorhombic distortion. First of all, the orthorhombic phase was not reported for FeTe$_{1-y}$Se$_{y}$ (y$\sim$0.33) \cite{Horigane2009,Prokes2015}. Moreover, the same FS was recorded for different specimens (for a number of measured samples two possible orientations would be observed, if the orthorhombic distortion took place). Finally, in the case of orthorhombic distortion the signal usually originates from two domain orientations in a usually twinned crystal, and ARPES spectra would be a superposition od these.

The ARPES data are in qualitative agreement with the results of DFT calculations, which predicted that Co or Ni doping in FeSe results in considerable expansion of electron FS volume and reduction of hole FS volume.\cite{Ciechan2013} The different shape of electron pockets at $\overline{M}$ for Ni and Co doping is a symptom of non-rigid band shift in the studied system. The nature of this difference is not understood so far.

\begin{figure*}[!]
\centering\includegraphics[width=\textwidth]{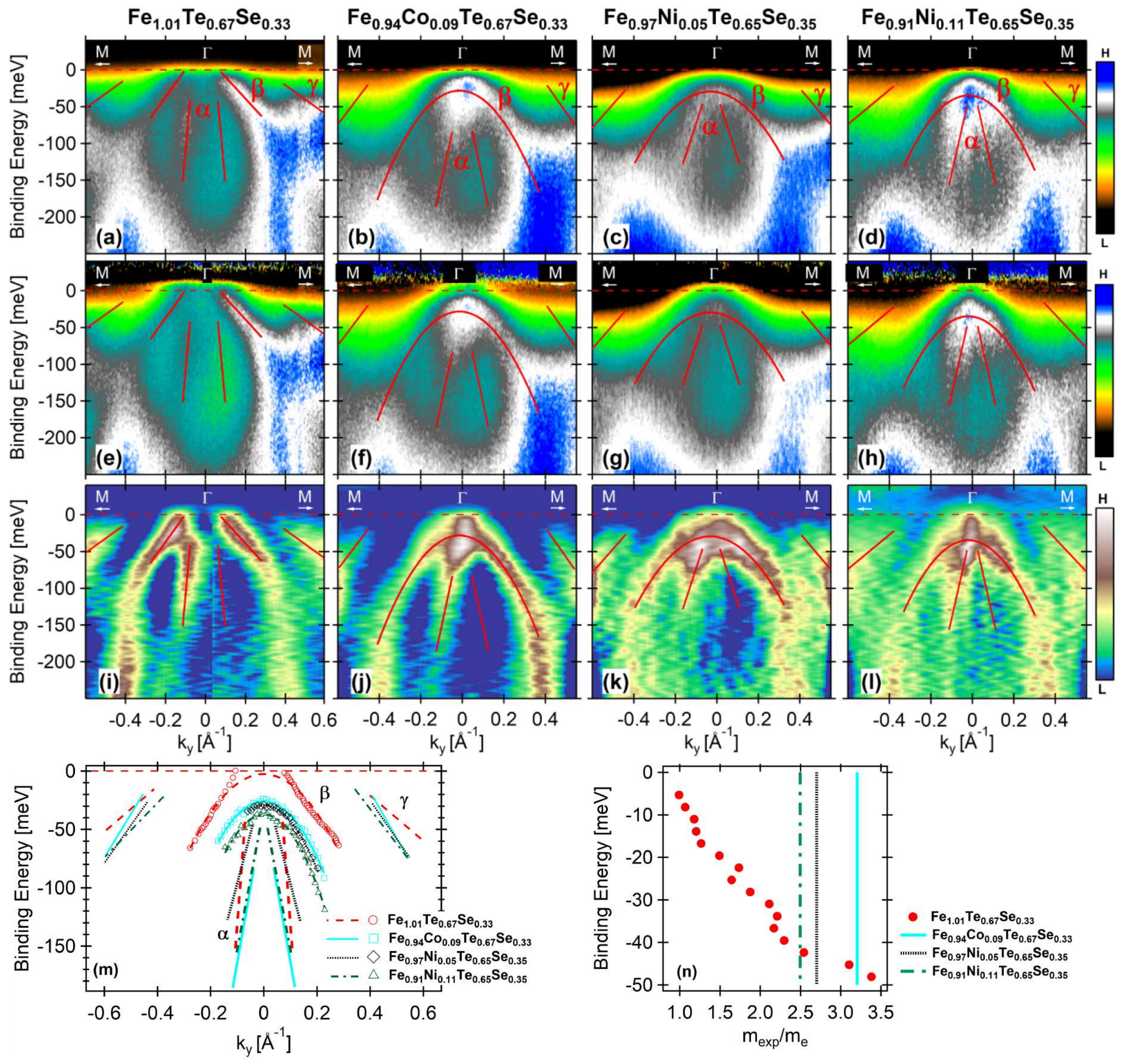}
\caption{Band structure of Fe$_{1-x}$M$_{x}$Te$_{1-y}$Se$_{y}$ (M=Co, Ni, y$\sim$0.35) systems in the region of $\overline{\Gamma}$ point obtained by ARPES at T=18 K. The measurements were performed in $\overline{M}-\overline{\Gamma}-\overline{M}$ direction along paths marked with "1" in [Fig 1 (c-f)]. Raw data: (a) Fe$_{1.01}$Te$_{0.67}$Se$_{0.33}$, (b) Fe$_{0.94}$Co$_{0.09}$Te$_{0.67}$Se$_{0.33}$, (c) Fe$_{0.97}$Ni$_{0.05}$Te$_{0.65}$Se$_{0.35}$, (d) Fe$_{0.91}$Ni$_{0.11}$Te$_{0.65}$Se$_{0.35}$. (e-h) – spectra from (a-d) normalized by the Fermi-Dirac distribution, (i-l) – 2D curvature plots \cite{PZhangRSI2011} obtained for the spectra shown in (a-d). (m) Hole band dispersions. For $\beta$ hole bands experimental points were estimated from MDC and EDC fitting to the spectra. Parabolas obtained from MDCs and EDCs maxima fitting are shown. Dispersions of $\alpha$ and $\gamma$ bands are indicated by maxima in the 2D curvature plots. (n) Effective mass for $\beta$ band expressed in free electron masses. In the case of  Fe$_{1.01}$Te$_{0.67}$Se$_{0.33}$ effective mass appears to be energy dependent, for the doped systems parabolic dispersion yields constant  $m^{\star}$.}
\label{fig2}
\end{figure*}

The effect of doping on the region of the $\overline{\Gamma}$ point is visualized in the Fig.~\ref{fig2}, which presents ARPES spectra taken along the lines “1” ($\overline{M}-\overline{\Gamma}-\overline{M}$ direction) shown in the Fig.~\ref{fig1}~(c)-(f). Band structure along $\overline{M}-\overline{\Gamma}-\overline{M}$ for Fe$_{1.01}$Te$_{0.67}$Se$_{0.33}$ reveals three hole pockets at $\overline{\Gamma}$ (Fig.~\ref{fig2}(a)), which agrees with the previous ARPES results for this system. \cite{Chen2010,Tamai2010,Nakayama2010,Ieki2014} However, we have detected neither electron band nor Dirac cones, which have been observed in a system with higher Se content (FeTe$_{0.55}$Se$_{0.45}$) by laser ARPES study.\cite{PZhangScience2018} These features were well resolved in the experimental geometry corresponding to $\pi$ polarization.\cite{PZhangScience2018} The raw ARPES data for Co and Ni doped samples (Fig.~\ref{fig2}(b)-(d)) show only two inner hole pockets. Fermi-Dirac normalized spectra (Fig.~\ref{fig2}(e)-(h)) reveal the same pockets as the raw data. However, the calculated 2D curvature \cite{PZhangRSI2011} (Fig.~\ref{fig2}(i)-(l)) recovers three hole bands for all the investigated systems. Dispersions of $\beta$ pockets were estimated by means of fitting the Gauss function to both MDCs and energy distribution curves (EDCs) from the data normalized by the Fermi-Dirac distribution (Fig.~\ref{fig2}~(e)-(h)). In the case of the doped samples, parabolas were fitted to $\beta$ pockets (Fig.~\ref{fig2}(b)-(d), (f)-(h)). Dispersions of $\alpha$ and $\gamma$ pockets are presented as indicative lines corresponding to maxima at the 2D curvature plots. 

It is clear that $\beta$ band, which forms a part of FS of the undoped system, is shifted below the Fermi energy (E$_{F}$) for all the doped samples (Fig.~\ref{fig2}(m)). The situation of the $\alpha$ pocket is less clear. However, it is visible that electron doping moves this band down and changes the slope. For the undoped sample $\alpha$ band may participate in FS but according to the found dispersions it should also be removed by doping. The data indicate that $\gamma$ pocket survives the electron doping. The removal of $\beta$ and probably $\alpha$ pockets changes the topology of FS, what is interpreted as a Lifshitz transition. This sheds a new light on the peculiar behavior of~R$_{H}$ \cite{Bezusyy2015}, which first increases with Ni or Co doping and next becomes negative for low temperature data. The disappearance of $\beta$ and probably $\alpha$ hole pockets decreases the hole FS volume and reduces the number of holes in the system. In parallel, the volume of electron pockets, which is proportional to electron band filling, grows with doping. Such evolution of FS leads to negative~R$_{H}$. In fact, R$_{H}$ for Fe$_{0.94}$Co$_{0.09}$Te$_{0.67}$Se$_{0.33}$ and Fe$_{0.97}$Ni$_{0.05}$Te$_{0.65}$Se$_{0.35}$ is still positive for all temperatures but these systems are located close to the dopant concentration, for which low temperature R$_{H}$ changes sign. It is already negative for Fe$_{0.91}$Ni$_{0.11}$Te$_{0.65}$Se$_{0.35}$ at low temperature. ARPES data also explain the fact that R$_{H}$ remains positive for all doping levels at room temperature. The tops of $\beta$ pocket are located at 26, 29, and 37~meV binding energy for Fe$_{0.94}$Co$_{0.09}$Te$_{0.67}$Se$_{0.33}$, Fe$_{0.97}$Ni$_{0.05}$Te$_{0.65}$Se$_{0.35}$ and Fe$_{0.91}$Ni$_{0.11}$Te$_{0.65}$Se$_{0.35}$, respectively. At room temperature such binding energies are accessible for thermal scattering, what includes $\beta$ band in FS effectively, delivering more hole carriers. 

It is noteworthy that the $\beta$ band has a parabolic shape, when it does not participate in FS. For Fe$_{1.01}$Te$_{0.67}$Se$_{0.33}$ its dispersion near E$_{F}$ is strongly modified. The effective mass calculated with formula $m(E)=\hbar^2 k(E) {dk}/{dE}$ (Fig.~\ref{fig2}(n)) appears to be strongly reduced with lowering energy, whereas it is constant for parabolic dispersions of the doped systems. This visualizes correlation effects, which strongly influence the dispersion near E$_{F}$. Previously, qualitatively similar effect was observed for NaFe$_{0.978}$Co$_{0.022}$As,\cite{Charnukha2015} and it was attributed to orbital mixing caused by strong spin-orbit coupling.

\begin{figure*}[!]
\includegraphics[width=0.8165\textwidth]{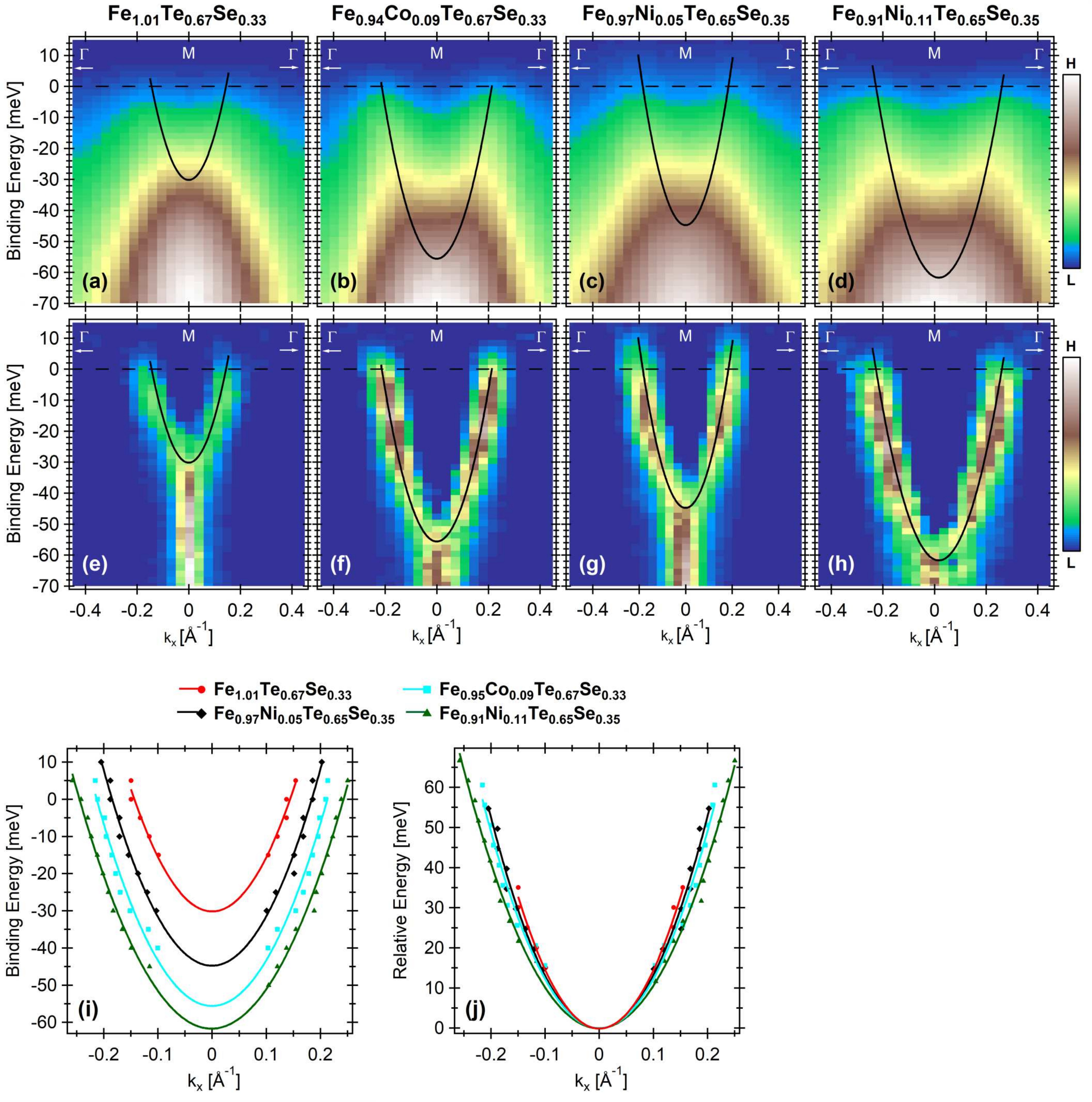}
\caption{Band structure of Fe$_{1-x}$M$_{x}$Te$_{1-y}$Se$_{y}$ (M=Co, Ni, y$\sim$0.35) systems in the region of $\overline{M}$ point obtained by ARPES at T=18~K. The measurements were performed along paths marked with "2" in [Fig 1 c-f]. Raw data are shown for: (a) Fe$_{1.01}$Te$_{0.67}$Se$_{0.33}$, (b) Fe$_{0.94}$Co$_{0.09}$Te$_{0.67}$Se$_{0.33}$, (c) Fe$_{0.97}$Ni$_{0.05}$Te$_{0.65}$Se$_{0.35}$, (d) Fe$_{0.91}$Ni$_{0.11}$Te$_{0.65}$Se$_{0.35}$. (e-h) present 2D curvature plots \cite{PZhangRSI2011} corresponding to data from (a-d). Black solid lines represent parabolas fitted to maxima of MDCs. The fitted parabolic dispersions with MDCs maxima are shown in (i).  The same dispersions with shifts in energy leading to overlap of the bottom of the band (j).}
\label{fig3}
\end{figure*}

The electron pockets observed along lines “2” from Fig.~\ref{fig1} are shown in the Fig.~\ref{fig3}. Only one pocket is resolved in both the raw data (Fig.~\ref{fig3}(a)-(d)) and 2D curvature (Fig.~\ref{fig3}(e)-(h)). The dispersion was determined by fitting MDCs, it also agrees with the maxima at 2D curvature. Electron bands follow well parabolas with no considerable correlation effects, which would be reflected in a more complex shape. The changes of the binding energies for both electron and $\beta$ hole bands as well as corresponding effective masses are presented in Tab.~\ref{tab1}. The binding energy shifts of hole and electron pockets are similar for Fe$_{0.94}$Co$_{0.09}$Te$_{0.67}$Se$_{0.33}$, and Fe$_{0.91}$Ni$_{0.11}$Te$_{0.65}$Se$_{0.35}$, what would roughly indicate a rigid band shift. The shifts determined for Fe$_{0.97}$Ni$_{0.05}$Te$_{0.65}$Se$_{0.35}$ contradict this thesis.

\begin{table}[t]
\setlength{\extrarowheight}{2pt}
\begin{footnotesize}
\caption{Energy shifts of $\beta$ hole pockets and electron pockets at $k_{y}=1.2~\si{\angstrom}^{-1}$ and corresponding effective masses expressed in free electron mass.}
\begin{tabular}{|c|c|c|c|c|}

\hline
\multirow{2}{1.25cm}{Compound} & \multicolumn{2}{c|}{$\beta$ Hole Pocket} & \multicolumn{2}{c|}{Electron Pocket}\\

\cline{2-5}
 &Band&Effective&Band &Effective\\
&Shift &Mass &Shift &Mass\\
&[meV]&[m$_e$] & [meV] &[m$_e$] \\
\hline

Fe$_{1.01}$Te$_{0.67}$Se$_{0.33}$ & 0 & - & 0 & 2.6(3)\\
Fe$_{0.94}$Co$_{0.09}$Te$_{0.67}$Se$_{0.33}$  & 23(1)  & 3.2(2) & 25(3)  & 3.1(2)\\
Fe$_{0.97}$Ni$_{0.05}$Te$_{0.65}$Se$_{0.35}$ & 26(1)  & 2.7(1) & 14(3)  & 2.8(2)\\
Fe$_{0.91}$Ni$_{0.11}$Te$_{0.65}$Se$_{0.35}$ & 35(1)  & 2.5(1) & 32(3)  & 3.7(1)\\ 

\hline
\end{tabular}
\label{tab1}
\end{footnotesize}
\end{table}

Effective masses ($m^{\star}$) can be easily estimated for bands with a parabolic shape (Tab.~\ref{tab1}). The $\beta$ band at $\overline{\Gamma}$ point allows to determine the effective mass to be between 3.2~m$_{e}$ and 2.5~m$_{e}$. We do not show $m^{\star}$ for the undoped sample, which is not constant but decreases with energy. For the electron band at the $\overline{M}$ point $m^{\star}$ grows with the amount of electrons delivered by Co or Ni, which means that correlations are enhanced there with electron doping.

\begin{figure*}[!]
\includegraphics[width=\textwidth]{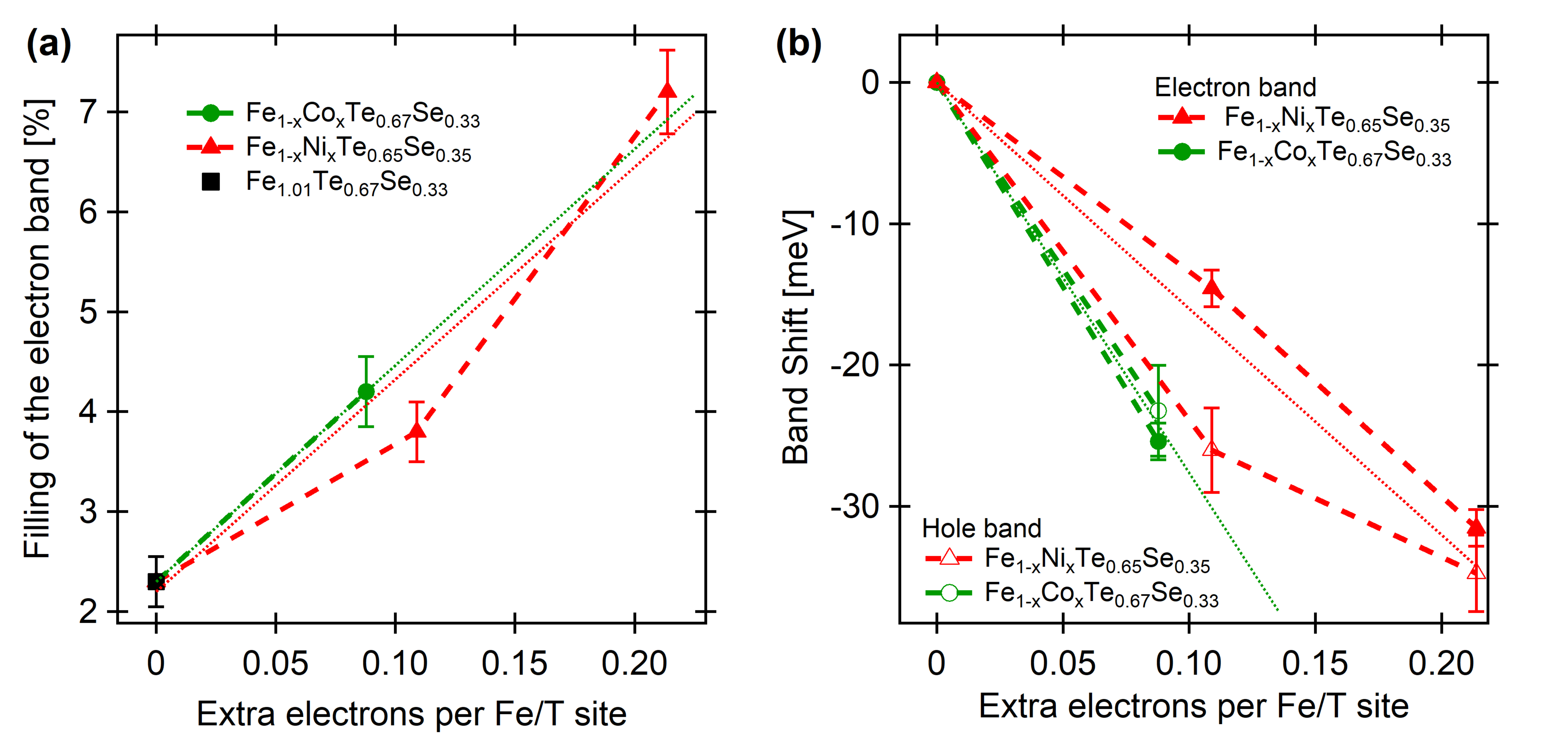}
\caption{Effect of doping on band structure of Fe$_{1-x}$M$_{x}$Te$_{1-y}$Se$_{y}$ (M=Co, Ni, y$\sim$0.35) estimated from ARPES as a function of a nominal amount of doped electrons assuming that Co and Ni donate 1 and 2 extra electrons, respectively. (a) Filling of the electron band determined from the area of solid black ovals in [Fig. 1 (c-f)]. (b) Band shifts estimated from the electron band [Fig. 3] and $\beta$ hole band [Fig 2] dispersions. The dotted lines show average effect of doping (fitted line), the dashed lines connect points.}
\label{fig4}
\end{figure*}

Co and Ni doping has a direct effect on band filling. The increase of FS volume for the electron pocket is given in the Fig.~\ref{fig4}(a). The results in this figure are normalized per doped electron, i.e. we assume that Co donates one extra electron and Ni delivers two. If we postulate linear trends to compare effects of doping quantitatively, the increase of number of states filled by electrons is 5$\%$ higher for Co substitution. Band shifts (Fig.~\ref{fig4}(b)) resulting from doping seem to be larger for Co and by assuming linear dependencies we can estimate that efficiency of transferring electrons is about 40$\%$ higher when compared to Ni. {However, this is FS volume, which is more directly related to a number of doped carriers. The shape of electron pocket, which is circular or elliptical depending on a dopant points to a non-rigid band shift in the system. In such a case it is more difficult to compare band shifts for Ni and Co doped compounds. Despite a relatively lower effect of doping} Ni causes stronger suppression of superconductivity and more pronounced effect on transport properties. \cite{Bezusyy2015} First of all, more efficient impurity scattering was predicted for Ni.\cite{PZhangRSI2011} Moreover, it is known that Co and Ni doping may have a different influence on the correlations in the system,  \cite{Vilmercati2016, Thirupathaiah2016} which have impact on superconductivity as well. Our data, which reveal less effective electron doping of Ni, confirm that stronger correlation effects must take place in samples with this element. In fact, more efficient impurity scattering is predicted for Ni.\cite{Herbig2016}  Fe$_{1-x}$M$_{x}$Te$_{1-y}$Se$_{y}$ (y$\sim$0.35) may be considered as a two-dimensional system to some extent. For a purely two-dimensional system with weak correlations i.e. with parabolic dispersions in 2-dimensions a rigid band shift should result in a proportional increase of band filling and band shift. This assumption is not convergent with our results. Finally, a rigid band shift predicted by DFT calculations for 10$\%$ of Co or 5$\%$ of Ni in FeSe \cite{Ciechan2013} should amount to 70~meV, which is much larger than the value of $\sim$ 25~meV found in our experiment. 

\begin{figure*}[!]
\includegraphics[width=0.75\textwidth]{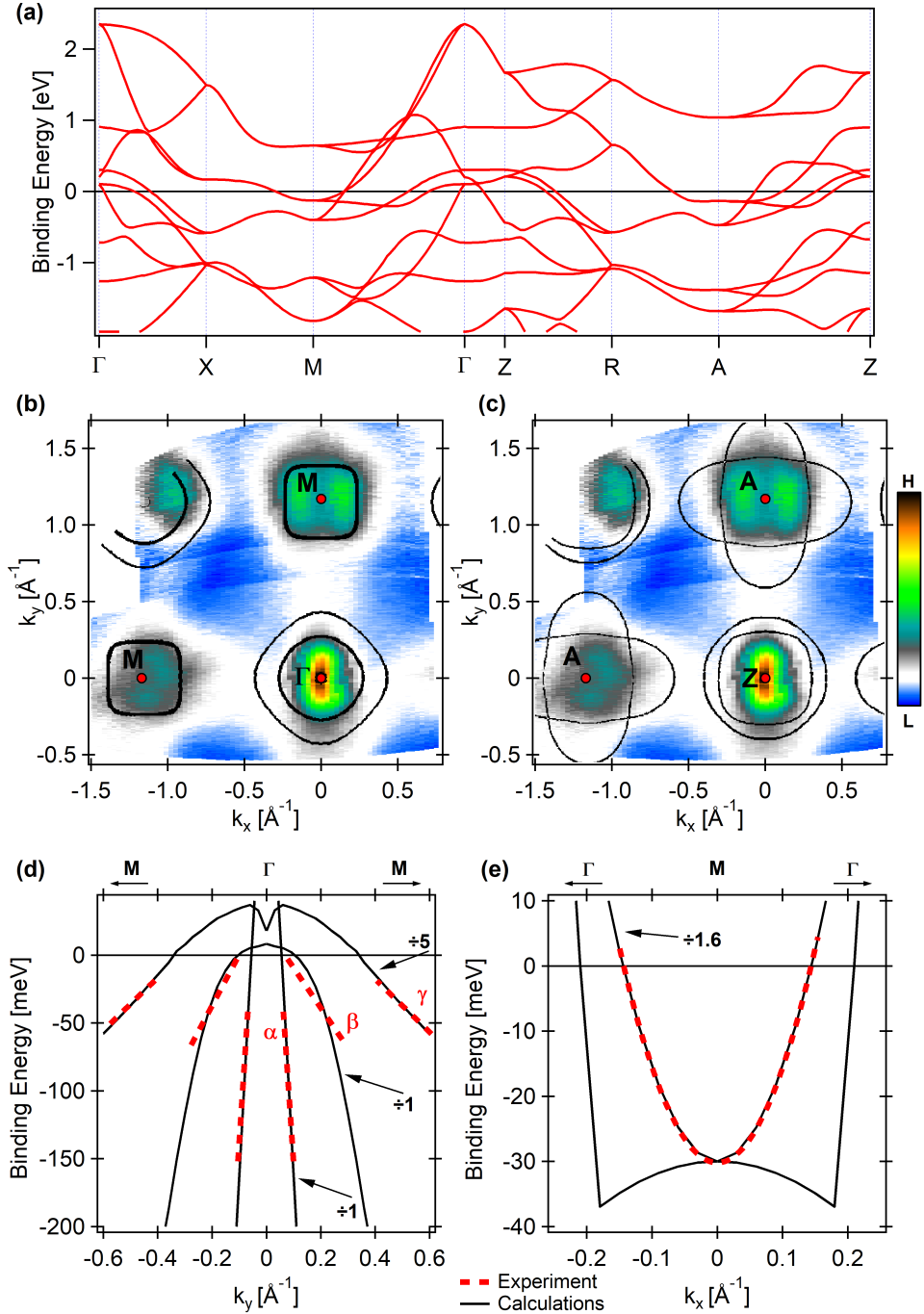}
\caption{(a) Band structure of FeTe obtained with APW-lo calculations. The theoretical Fermi surface (black solid lines) along (b) $\Gamma$-M-X plane and (c) Z-A-R plane is superimposed on experimental FS map for Fe$_{1.01}$Te$_{0.67}$Se$_{0.33}$. (d,e) Comparison between theoretical (black solid lines) and experimental (red dotted lines) dispersions along M-$\Gamma$-M direction at (d) $\Gamma$ and (e) M point. Theoretical binding energies for $\alpha$, $\beta$ and $\gamma$ bands were shifted by 70~meV, 92~meV and -110~meV and divided by~1, 1 and 5, respectively. The theoretical binding energy for electron pocket (e) had to be divided by 1.6 after being shifted by 80~meV.}
\label{fig5}
\end{figure*}

The experimental data for the undoped Fe$_{1.01}$Te$_{0.67}$Se$_{0.33}$ can be compared to APW+lo calculations performed for stoichiometric FeTe (Fig.~\ref{fig5}). The calculated FSs are shown along $\Gamma$-M-X or Z-A-R planes. In both cases theoretical FSs have much larger volume as compared to experimental ones. In other words the results indicate shrinking of FS (Fig.~\ref{fig5} (a),(b)), which was explained before as originating from interband scattering. \cite{Ortenzi2009} It appears that calculations performed along M-$\Gamma$-M are more consistent with experimental data for $\overline{M}-\overline{\Gamma}-\overline{M}$ than the ones obtained along A-Z-A. This indicates that experimental band structure collected at $h\nu=21.218$~eV  corresponds rather to $\Gamma$ point crossing (or its vicinity) at normal emission. Theoretical dispersions are compared with the experimental ones, with the renormalization factor of 1, 1 and 5, as well as energy shifts of 70~meV, 92~meV, and -110~meV for $\alpha$, $\beta$ and $\gamma$ pockets, respectively. These results are in qualitative agreement with previous study for FeTe$_{0.5}$Se$_{0.5}$,\cite{Thirupathaiah2016} where also $\gamma$ pocket appeared to be highly renormalized. The electron pocket observed at $\overline{M}$ point as with theory for the renormalization factor of 1.6 with energy shift of 0.08~eV (Fig.~\ref{fig5}(d)).

FS topology should be analyzed in terms of possible scattering between hole and electron pockets. For the superconducting Fe$_{1.01}$Te$_{0.67}$Se$_{0.33}$ system the radius of  $\beta$ and $\gamma$ hole pockets expressed in $\Gamma$-M distance is roughly estimated as~0.075 and 0.23, respectively. The $\alpha$ pocket is rather smaller than $\beta$. On the other hand the radius of the estimated electron pocket at $\overline{M}$ amounts to 0.11 in the same units. Hence, the difference of radii between the electron pocket and either $\beta$ or $\gamma$ hole pockets is $\sim0.037$ and 0.12, respectively. As a result of Co/Ni doping these values will grow. Then, it is concluded that for superconducting Fe$_{1.01}$Te$_{0.67}$Se$_{0.33}$ there is a considerable deviation from perfect nesting due to the mismatch between electron and hole pockets. Earlier, it was proposed that such a marginal breakdown of nesting will support the appearance of spin fluctuations as a perfect nesting would rather induce a spin density wave.\cite{Lee2012} Here, we can compare our results to corresponding mismatch in nesting obtained by DMFT and ARPES for (Li,Na)FeAs.\cite{Lee2012} The digitalization of the data from Fig.~2(c) from Ref.~\onlinecite{Lee2012}  yielded that average difference in radii between hole and corresponding (e.g. with the lowest mismatch) electron pocket is 0.08 and 0.10~$\Gamma$-M distance for $\beta$ and  $\gamma$, respectively. The first value is quite larger than in the case of Fe$_{1.01}$Te$_{0.67}$Se$_{0.33}$. However, a general rule that imperfect nesting supports superconductivity should also hold for Fe-11 systems.

It is interesting to compare the effect of Co doping on band structure of a superconductor from Fe-122 family, namely Ba(Fe$_{1-x}$Co$_{x}$)$_{2}$As$_{2}$ \cite{Liu2010, Liu2011, Brouet2009, Vilmercati2016} and Fe$_{1-x}$Co$_{x}$Te$_{0.67}$Se$_{0.33}$ from this study, which represents Fe-11 systems. Increasing amount of Co in Ba(Fe$_{1-x}$Co$_{x}$)$_{2}$As$_{2}$ first destroys magnetic order, next induces superconductivity, which is further suppressed in the overdoped region. In contrast, the only role of Co in Fe$_{1-x}$Co$_{x}$Te$_{0.67}$Se$_{0.33}$ is a suppression of superconductivity. The FS of underdoped Ba(Fe$_{1-x}$Co$_{x}$)$_{2}$As$_{2}$ contains characteristic “flower petals” around the M point (according to convention in the quoted paper it is denoted as X) in the antiferromagnetic phase. These structures are absent for Fe$_{1-x}$Co$_{x}$Te$_{0.67}$Se$_{0.33}$. However, for higher Co content in Fe-122 the analogy to Fe-11 is found. FS of the optimally doped Ba(Fe$_{1-x}$Co$_{x}$)$_{2}$As$_{2}$ is analogous to FeTe$_{1-y}$Se$_{y}$ (y$\sim$0.35). In both cases hole and electron pockets have similar volumes and the situation close to nesting takes place for these parts of FS. This imperfect nesting is removed with further Co doping in Fe-122 and Fe-11. In both highly doped systems FS with large electron pockets at $\overline{M}$ and small or removed pockets at $\Gamma$ has no more nesting properties. It should be stressed that nesting is removed faster in Ni doped Fe$_{1-x}$Ni$_{x}$Te$_{0.65}$Se$_{0.35}$ samples, where electron pockets become elliptical. It is noteworthy, that we did not observe electron pockets, which appear at $\Gamma$ in highly doped Ba(Fe$_{1-x}$Co$_{x}$)$_{2}$As$_{2}$.\cite{Liu2010, Liu2011}

\section{Conclusions}

ARPES studies revealed clear evidence of electron doping on band structure of FeTe$_{1-y}$Se$_{y}$ (y$\sim$0.35) superconductor, what is an effect of Co or Ni substitution for iron. This is reflected in comparable shifts of both hole and electron pockets. The electron FS volume increases, while a part of hole FS disappears, which is a realization of Lifshitz transition. The efficiency of electron doping is higher for Co; with 5\% higher rate of filling electron pocket and 40\% larger band shifts, if one assumes that Ni should yield twice more carriers than Co. Therefore, more remarkable  influence of Ni on transport properties and significant suppression of superconductivity observed before \cite{Bezusyy2015} must be definitely attributed to strong correlation effects and not the contribution of carrier doping. The elliptical shape of electron pockets appears as an effect of Ni doping, while it is not observed for Co substitution. This fact together with effective mass growth at $\overline{M}$ point with doping, considerable reduction of $m^{\star}$ in the undoped $\beta$ hole pocket near E$_{F}$ and large difference in band shifts for comparable increase of band fillings for Co and Ni dopant indicates that there are deviations from rigid band shift scenario. While a clear departure from perfect FS nesting between hole and electron pockets is observed for FeTe$_{1-y}$Se$_{y}$ (y$\sim$0.35) superconductor, sizes of these pockets diverge further with Co or Ni doping. Such FS evolution is analogous to overdoped Ba(Fe$_{1-x}$Co$_{x}$)$_{2}$As$_{2}$, which may indicate that Co/Ni doped Fe-11 should be considered as overdoped system.

\section{Acknowledgments}
Support of the Polish Ministry of Science and Higher Education under the grant 7150/E-338/M/2018 is acknowledged. The work has been supported by Polish NSC Grant No. 2014/15/B/ST3/03889.

\end{document}